\RequirePackage{fix-cm}
\documentclass[smallextended,natbib]{svjour3}       
\smartqed  
\usepackage{graphicx}
\usepackage{mathptmx}      
\usepackage[usenames,dvipsnames]{xcolor}
\usepackage[breaklinks = true]{hyperref}
\hypersetup{
    colorlinks,
    linkcolor={blue},
    citecolor={blue},
    urlcolor={blue}
}
\usepackage{breakcites}
\usepackage{amssymb}
\usepackage{amsmath}
\usepackage{blkarray}
\usepackage{multicol}
\journalname{Experimental Astronomy}
\begin{document}

\title{On the use of Logistic Regression for stellar classification\thanks{This work has been partly
funded by the Ministry of Economy, Industry and Competitiveness of Spain 
through grants ESP2014-54243-R, ESP2015-68908-R and TIN2015-66471-P
as well as by the local Government of Madrid through grant S2013/ICE2845.}
}
\subtitle{An application to colour-colour diagrams}


\author{Leire Beitia-Antero         \and
  Javier Y\'a\~nez \and
  Ana I. G\'omez de Castro
}

\authorrunning{Beitia-Antero et al.} 

\institute{Facultad de Matem\'aticas, Universidad Complutense de
Madrid, Plaza de las Ciencias 3, 28040 Madrid, Spain   \and AEGORA
Research Group, Universidad Complutense de Madrid, Plaza de las
Ciencias 3, 28040 Madrid, Spain \email{aig@ucm.es}  
}

\date{Received: XXXX / Accepted: YYYY}

\maketitle

\begin{abstract}
We are totally immersed in the Big Data era and
reliable algorithms and methods for data classification
are instrumental for astronomical research.  Random Forest and
Support Vector Machines algorithms have become popular over 
the last few years and they are widely used for different stellar
classification problems. In this article, we explore an alternative
supervised classification method scarcely exploited in astronomy, Logistic Regression, that has been applied successfully
in other scientific areas, particularly
biostatistics. We have applied this method in order to derive membership probabilities for potential T Tauri
  star candidates  from ultraviolet-infrared colour-colour diagrams.

\end{abstract}

\section{Introduction}

Stellar classification  has evolved in leaps and bounds for the
last several years. Nowadays, lots of stellar catalogues are
available online, with hundreds of thousands of sources, and
so machine learning algorithms have become a popular and powerful
tool to classify these enormous datasets. One of the commonly-used
supervised methods is Random Forest (\cite{Breiman2001}), that
has been applied in order to classify variable stars from
the Hipparcos sample (\cite{Dubath2011, Johnston2017, Pichara2013}) or
Kepler catalogue (\cite{Bass2016}).
Another widely used automatic classification algorithm is Support Vector Machines
(\cite{Cortes1995, Vapnik1999, Cristianini2000})
that has been applied by \citet{Kurcz2016} to the WISE
catalogue to obtain star, galaxy and quasar catalogues, and 
it is also the main algorithm used to classify all the Gaia sources
(\cite{BellasVelidis2012}). There exist other algorithms based on bayesian methods
such as those proposed by \cite{Picaud2005} or \cite{BailerJones2011}
that allow to improve the accuracy of the classification.
All these methods are designed to cope with huge amounts of
sources and are suited for data with multidimensional nature. However,
in some cases the simplicity of the problem or a reduced sample
are not worth the effort.\par

The logistic function was introduced by Verhulst (\cite{Verhulst1845, Verhulst1847}) in the $19^{\rm th}$ century
to describe population growth in biological experiments  and 
was early applied on chemical processes (\cite{Reed1929}); a later
formulation was given by \cite{Cox1958}. Currently, Logistic
Regression is one of the most used techniques in Biostatistics (see
\cite{Hosmer2013}).  Logistic Regression is a
multivariate analysis model that predicts the probability of membership to any
class based on the values of some predictor variables; these
variables are not constrained to follow a given (normal) distribution, not
even be continuous. Although this method is not widely used in astrophysics, 
it has been recently applied by \cite{Huppenkothen2017}
in order to study the variablility of the galactic black hole binary
GRS 1915+105. Moreover, a variation of the method (Boosted Logistic
Regression) has been compared with Random Forest by \cite{SazP2016}
to classify gamma-ray sources into two groups and they have shown that
similar accuracy is achieved with both methods.\par

The Galaxy Evolution Explorer (GALEX) mission (\cite{Martin2005}) has surveyed the sky
in two ultraviolet bands, the far ultraviolet (FUV) band (1344~\AA- 1786~\AA )
and the near ultraviolet (NUV) band (1771~\AA - 2831~\AA ). Already
decommissioned,
during its life time GALEX detected about 200,000 million sources
(\cite{Bianchi2014})
that together with the 2MASS survey (\cite{Skrutskie2006}) provide an
incredible wealth of information to study 
star formation and the interstellar medium in the Galaxy. Some applications to the
Taurus  and Orion star forming regions can be found in
\cite{GdC2015,GdC15,Sanchez2014,Beitia2017}.\par

In \cite{GdC2015} (hereafter GdC2015) several colour-colour diagrams 
were used to identify T Tauri Stars (TTSs) from the background stellar population. Although the classification method was purely empirical, some
areas were identified where reliable candidates should be located but it
 neither 
assigned an uncertainty nor a probability of membership to a given class of the candidates.
This is a serious drawback to elaborate sensible lists of candidates for
subsequent observation.
Logistic Regression algorithms model robustly the probability
of membership, providing
a  measure of the classification errors.\par 

In this work, we apply Logistic Regression to the empirical classification done in GdC2015
and show how membership probabilities can be calculated.
In Section \ref{sec-logistics}, the method is described and in Section \ref{sec-TT} is applied
to the sample in GdC2015. In Section \ref{sec-MSclas} results of the application
of the algorithm to a sample of Main Sequence stars are shown and the
computational details are included in the Appendix.
A short summary of the results is provided in 
Sec. \ref{sec-conclusions}.

\section{Logistic Regression }\label{sec-logistics}

Logistic Regression is a machine learning model used to solve supervised classification problems:
  given a sample of objects characterised by the values of some predictor or independent
  variables, and knowing the class (dependent variable) of each one, the variables are used
  to classify the objects in the given groups. The solution of the Logistic Regression model will be
  a function that, given a new set of values of dependent variables, will compute the probability of belonging
to any of the pre-defined classes. \par

Mathematically, (see for instance \cite{Hastie2013} or \cite{Pradhan2010}), 
the probability of occurrence of some event, in
our case the probability that an object belongs to a particular
class or category, is modelled as:

\begin{equation}
\label{eq-logistic-1} p= \begin{array}{c} 1 \\ \hline 1 + exp(z)
\end{array}
\end{equation}

\noindent where $z$ is a linear combination of some predictor
variables:

\begin{equation}
\label{eq-logistic-2} z = \beta_0 + \beta_1 x_1 + \beta_2 x_2 +
\cdots + \beta_n x_n
\end{equation}

The probability $p$ varies from $0$ to $1$ on an $S$-shaped curve,
$\beta_0$ is the intercept of the model and $\beta_i$
($i=1,2,\ldots,n$) are the coefficients of the Logistic
Regression model. An appropriate selection of the predictor variables $x_1,x_2,\ldots,x_n$ is
crucial for an accurate modelling and any combination of the
original variables is allowed by the method: $x_j = x_1^2; x_k = x_1 x_2; ...$. \par

Hence, the classification problem can be expressed as,
\emph{Given a set of classes $\{ 1,2,\ldots,K \}$, and given an astronomical object $o
\in \cal{O}$ characterised by a a set of variables ${\bf
x}^o=(x_1^o,x_2^o,\ldots,x_n^o)$, what is $P(G=k|{\bf x}^o)$, the
probability that the object belongs to any of these classes?}

 The answer will distinguish between the probability of classes $1,2,\ldots,K-1$  and is given as,

\begin{equation}
\begin{array}{l}
P(G=k|{\bf x}^o) = \begin{array}{c} exp(\beta_{k0}+\sum_{i=1}^n \beta_{ki} x_i^o) \\
\hline 1 + \sum_{h=1}^{K-1} exp(\beta_{h0}+\sum_{i=1}^n \beta_{hi} x_i^o)
\end{array} \quad k \in \{1,2,\ldots,K-1 \} \\  
\end{array}
\label{eq-logistic-3}
\end{equation}

\noindent
and the probability of class $K$,
\begin{equation}
\begin{array}{l}
P(G=K|{\bf x}^o) = \begin{array}{c} 1 \\
\hline 1 + \sum_{h=1}^{K-1} exp(\beta_{h0}+\sum_{i=1}^n \beta_{hi} x_i^o)
\end{array}
\end{array}
\label{eq-logistic-4}
\end{equation}

\noindent
By definition $\sum_{h=1}^{K}P(G=h|\textbf{x}^{o}) = 1, \forall \textbf{x}^{o}$. 

The parameters $\{\beta_{k,i},~ k=1,\ldots, K,~ i=1, \ldots, n\}$ are estimated from
a qualification sample $\{(x_{1}^{j}, \ldots,x_{n}^{j}), y^{j}$ $j=1,\ldots,m \}$
where $(x_{1}^{j}, \cdots,x_{n}^{j})$ are the variables that characterise
objects \textit{j}, $j = 1,\ldots,m$  respectively, and $y^{j} \in \{1,\ldots,K\}$  are  
the classes. 

In this type of problems,  the \emph{Classification Table} 
or \textit{Confusion Matrix} is commonly used. This is a square matrix $C$ with
dimension $K$, the number of classes, and such that any 
element $ c_{i,j}$ indicates the number of objects of class $i$
that are classified by the model into class $j$, for all $i,j \in \{1,2,\ldots,K \}$
(see Table \ref{tab-claK}).

\begin{table}[h!]
\centering
\caption{Classification Table (Confusion Matrix) for $K$ classes}
\label{tab-claK}      
\begin{tabular}{r|rrrr} \hline 

& \multicolumn {4}{c}{Predicted}\\ \cline{2-5} \\

Classified &{\it Class 1} & {\it Class 2} & $\cdots$ & {\it Class $K$}\\
\hline \noalign{\smallskip}
{\it Class 1} & $c_{1,1}$ & $c_{1,2}$ & $\cdots$ & $c_{1,K}$\\
{\it Class 2} & $c_{2,1}$ & $c_{2,2}$ & $\cdots$ & $c_{2,K}$\\
$\vdots$ & $\vdots$  & $\vdots$  & $\ddots$ & $\vdots$ \\
{\it Class $K$} & $c_{K,1}$ & $c_{K,2}$ & $\cdots$ & $c_{K,K}$ \\

\noalign{\smallskip}\hline

\end{tabular}
\end{table}

\section{Classification of T Tauri stars }\label{sec-TT}

Our scope is to use the Logistic Regression model to assign
a given star with known FUV, NUV, J and K magnitudes a probability of
being either a classical or weak-line TTS. GdC2015 obtained a qualification sample of 47 TTSs
detected in the Taurus molecular cloud that had been 
observed either by the International Ultraviolet Explorer (\cite{GdC1997}) or by  GALEX and that have a 2MASS counterpart; that sample
consists on 16 weak line TTSs (WTTSs) and 31 Classical TTSs (CTTSs). 
As pointed out by GdC2015, though WTTs are close to the main
  sequence and it is difficult to identify them directly from the diagram,
  a regression line was found; this was not the case of the CTTSs, that were
  sparsely distributed over an area far from the main sequence. In addition,
both groups seem to be mixed. For this reason, the crucial issue in this work is to define
properly the location of the Main Sequence in the FUV-NUV versus
J-K colour-colour diagram.

\subsection{Model 1: Main Sequence defined by the Field Stars in the area}

\begin{figure}
\centering
  \includegraphics[width = \columnwidth]{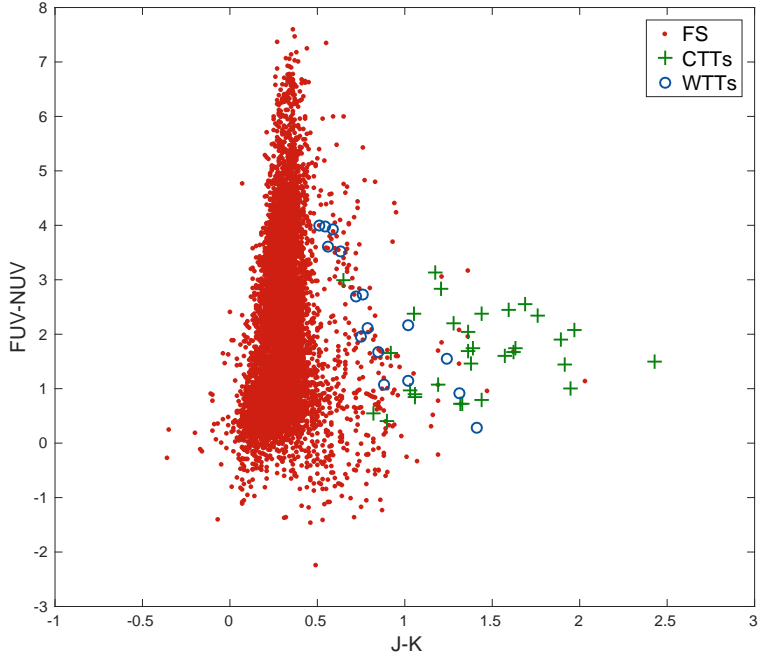}
  \caption{Initial sample of GALEX field stars (red dots),
    CTTs (green pluses) and WTTs (blue circles)
  from GdC2015.}
\label{fig:tt_sample}      
\end{figure}

In a first attempt, we used
a sample of 7348 stars observed by GALEX in Taurus, with both
  FUV and NUV magnitudes, that also had a 2MASS counterpart (see
  GdC2015 for details); this plain
  sample will be referred to as Field Stars (FS) and is shown in
Fig. \ref{fig:tt_sample} (red dots). Note that the separation
  between the WTTs and FS is rather fuzzy due to variations on the extinction
  values of some FS, and will complicate the differentiation of both classes. Besides,
  the size of the FS sample is enormous compared to the CTTs and WTTs ones and will
bias the results.

The problem will consist on discerning the probability that
  any of the FS is a CTTS or WTTS (or neither of them, and so they will remain
  as FS).
  Following the Logistic Regression formulation (Sec. \ref{sec-logistics}),
  we will determine the membership probability 
  to any of the three classes: CTTS ($k=1$), WTTS ($k = 2$) and FS ($k = 3$),
where higher probability of belonging to the FS class means that the object is not likely to be a TT. The variables considered are $x_{1} = J-K$, $x_{2} = FUV-NUV$,
$x_{3} = x_{1}^{2}$, $x_{4}=x_{2}^{2}$ and  $x_{5}=x_{1}x_{2}$;
variables $x_{3}$, $x_{4}$ and $x_{5}$ are introduced due to the
nonlinear nature of the sample (Fig. \ref{fig:tt_sample}). The family
of parameters to be estimated is:
\begin{equation}
  \big\{ \beta_{ki} \ | \ k \in \{ 1,2 \}  ; i \in \{ 0,1,2,\ldots, 5 \} \big\}
  \label{eq-parameters}
\end{equation}
where $k=1$ corresponds to the first class, CTTs, and $k=2$ corresponds to the second class, WTTs. A computational application was programmed in MATLAB (\cite{MATLAB}) to estimate the parameters as well as to obtain the associated graphics. The results are given in Table \ref{tab:lr_tt_direct}.

\begin{table}
  \centering
  \caption{Slope coefficients of the quadratic Logistic Regression model for the TT sample. $k=1$ coefficients correspond to the CTTs population while $k=2$
  fits the WTTs population.}
  \label{tab:lr_tt_direct}
  \begin{tabular}{ccccccc}
    \hline
    & $\beta_{k,0}$ & $\beta_{k,1}$ & $\beta_{k,2}$ & $\beta_{k,3}$ & $\beta_{k,4}$ & $\beta_{k,5}$ \\
    \hline
    $k=1$ & -14.1806 & 15.6095 & 0.9570 & -4.3114 & -0.2846 & 0.4122 \\
    $k = 2$ & -24.6297 & 34.0705 & 4.9589 & -12.9391 & -0.5180 & -2.5460 \\
    \hline
    
  \end{tabular}
\end{table}

\begin{figure}
\centering
  \includegraphics[width = \columnwidth]{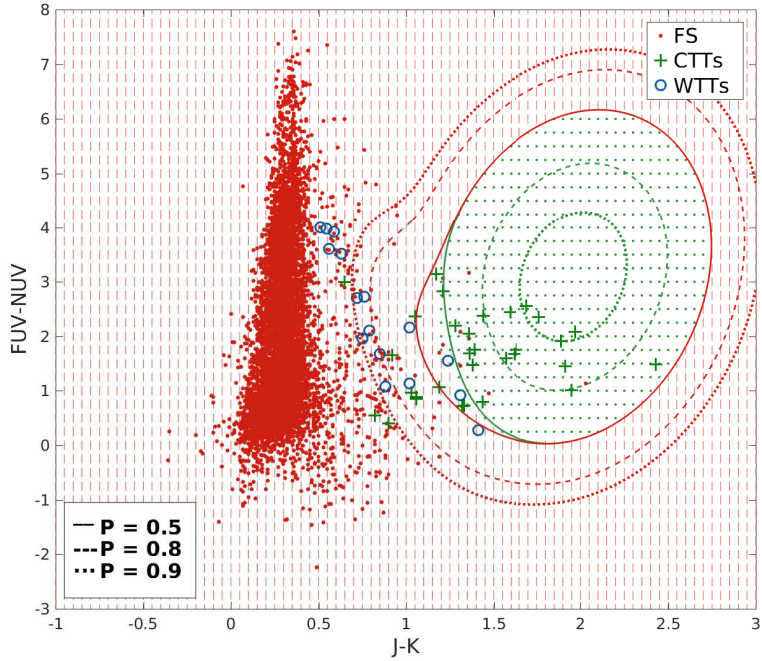}
  \caption{Results of the Logistic Regression fit of the FS and TTs sample from GdC2015. Each point
    of the plane has been assigned the most probable class with a color code: vertical red lines for FS and green points for CTTs. Also three probability lines have been plotted in order to separate the 0.5, 0.8 and 0.9
  probability regions for FS (red) and CTTs (green). The algorithm has ignored the WTTs class.}
\label{fig:tt_lr1}      
\end{figure}

 In Fig. \ref{fig:tt_lr1}, the most likely class for each point of the
 (J-K)-(FUV-NUV) plane has been determined from the Logistic Regression results
 (coloured mesh grid corresponds to the colour code of the sample: blue for WTTs,
   green for CTTs and red for FS). As occurs with every automated
   classification procedure, we find misclassified
   objects among our sample (mainly the WTTs population, that has been ignored) that should be minimum if the
   algorithm behaves properly. 
   The contours delimit the areas inside which the membership
   probability is higer to 0.5 (solid line), 0.8 (dashed line) and 0.9 (dotted
   line). Note that given the definition of the probabilities (Eqs. \ref{eq-logistic-3} and \ref{eq-logistic-4})
   and the fact that we are establishing three groups, the areas of, e.g. $P \geq 0.5$, may not be
   closed areas but rather open ones, as happens with the FS sample.
As mentioned above, virtually no point has been assigned to WTTs class. This is positively due to a
lack of balance in the sample as the most numerous population corresponds to FS whereas
the WTTs sample is the least populated. Moreover, FS are submitted to
different extinctions resulting in a broadening of the MS branch and as WTTSs lie pretty near of the FS population, 
the algorithm is obviously confusing both classes. The discouraging results of the confusion matrix
(see Table \ref{tab-model1}) support this fact.

\begin{table}[h!]
  \centering
\caption{Confusion Matrix for Model 1: initial unbalanced sample.}
\label{tab-model1}      
\begin{tabular}{r|rrr} \hline 

& \multicolumn {3}{c}{Predicted}\\ \cline{2-4}

Classified &{\it CTTS} & {\it WTTS} & {\it FS}\\
\hline \noalign{\smallskip}

{\it CTTS} & 19 & 0 & 12 \\
{\it WTTS} & 2 & 0 & 14\\
{\it FS} & 9 & 0 & 7339 \\

\noalign{\smallskip}\hline

\end{tabular}
\end{table}

Hence, it is necessary to compensate the sample in order to achieve better results.

\subsection{Model 2: Main sequence from a balanced sample of field stars. }
\begin{figure}
\centering
  \includegraphics[width = \columnwidth]{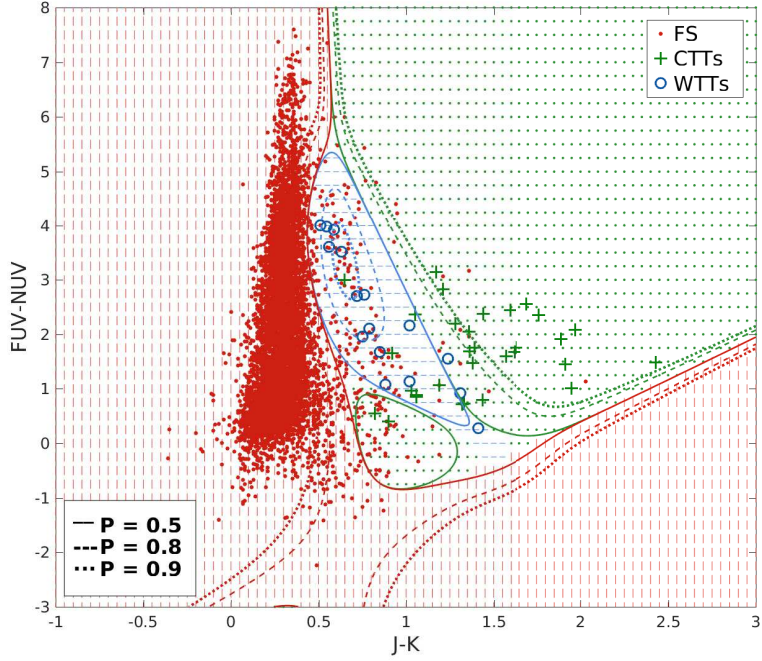}
  \caption{Results of the Logistic Regression fit with a sample balanced using the upper limits for
    the scaling coefficients, $\lambda_{1} = 237, \lambda_{2} = 459$. Some points
    of the plane has been assigned to the WTTs class (blue horizontal lines). Also three probability lines have been plotted in order to separate the 0.5, 0.8 and 0.9
  probability regions.}
\label{fig:tt_lr2}      
\end{figure}

As we will show below, the best solution is achieved by reducing the statistical weight of field stars. From a statistical point of view,
  it is sensible to have equal probability of belonging to any of the
  three classes (CTTS, WTTS or other, FS), supposing the sample is a simple random one. Given the
  fact that the FS population is much numerous than the TTs population, the algorithm could ignore the WTTs without
  a significant penalty on the results, as we have shown in the first approach. A possible solution for this problem is
to balance the CTTs and WTTs sample sizes by integer factors 
$\lambda_{1}$ and $\lambda_{2}$, respectively, in order to obtain more flexible
probability distributions that fit better to the spatial distribution of the sources. An upper limit for these parameters is established
by the size of the FS sample (7348), \textit{i.e.}, $\lambda_{1} = 7348/31 \sim 237$ and
$\lambda_{2} = 7348/16 \sim 459$. If
we perform again the Logistic Regression using these upper limits (see Table \ref{tab:lr_tt_upper}
for the slope coefficients), the results are 
highly improved as clear from Fig. \ref{fig:tt_lr2}, in the sense that the WTTs class has not been ignored and most
of the sources lie inside their $P>0.5$ probability area.

\begin{table}
  \centering
  \caption{Slope coefficients of the quadratic Logistic Regression model for the TT balanced sample
    using the upper limits for the scaling coefficients, $\lambda_{1} = 237, \lambda_{2} = 459$.}
  \label{tab:lr_tt_upper}
  \begin{tabular}{ccccccc}
    \hline
    & $\beta_{k,0}$ & $\beta_{k,1}$ & $\beta_{k,2}$ & $\beta_{k,3}$ & $\beta_{k,4}$ & $\beta_{k,5}$ \\
    \hline
    $k=1$   & -9.1677  & 17.8639 & -2.6685 & -7.2559  & 0.0526  & 4.4272 \\
    $k = 2$ & -24.9926 & 39.4334 & 6.2785  & -14.5633 & -0.7100 & -1.8300 \\
    \hline
    
  \end{tabular}
\end{table}

Formally, this improvement can be tracked through the Rate matrix, $R$.
$R$ is a  $K \times K$ matrix with elements $r_{i,j}$  defined as the 
fraction of the total sources belonging to class $i$ that
have been classified as class $j$, where $i,j \in \{ 1,2,\ldots, K
\}$ ($K=3$ in this case). This matrix is similar to the confusion matrix in that
the classification improves as the elements outside the main diagonal
decrease toward 0. The rate matrix for model 1 (see Sec.~3.1) is, 
\[
\begin{blockarray}{cccc}
  \begin{block}{c(ccc)}
    & 0.61 & 0.00 & 0.39\\
    \textit{R} = & 0.12 & 0.00 & 0.88\\
    & 0.00 & 0.00 & 1.00\\
  \end{block}
  \end{blockarray}
\]
that shows clearly that the FS class is overestimated. However, for the scaled model
considering $\lambda_{1} = 237, \lambda_{2} = 459$, the rate matrix is:

\[
\begin{blockarray}{cccc}
  \begin{block}{c(ccc)}
    & 0.74 & 0.26 & 0.00\\
    \textit{R} = & 0.12 & 0.88 & 0.00\\
    & 0.01 & 0.02 & 0.97\\
  \end{block}
  \end{blockarray}
\]

The case $R = I$, the identity matrix, will correspond to the
  perfect classification, where the predicted class of every source
  will be the real one. The difference between matrices $R$ and $I$ can
be considered a quality measure of the classification algorithm.
The final step is to determine 
$\lambda_{1}$ and $\lambda_{2}$ values in order to
improve as much as possible this rate matrix R. These integer values should be in the ranges $\lambda_{1} \in \{1,\cdots,237\}$
and $\lambda_{2} \in \{1,\cdots,459\}$. After several experiments, we
  have found that the best values in the sense of the matrix $R-I$ are
  $\lambda_{1}=69$, $\lambda_{2} = 86$, providing the confusion matrix
  showed in  table \ref{tab-balanced-sample} and rate matrix,

\[
\begin{blockarray}{cccc}
  \begin{block}{c(ccc)}
                 & 0.94 & 0.06 & 0.00\\
    \textit{R} = & 0.19 & 0.81 & 0.00\\
                 & 0.01 & 0.01 & 0.98\\
  \end{block}
  \end{blockarray}
\]

\begin{table}[h!]
  \centering
  \caption{Confusion Matrix for the model considering a balanced
    sample, with $\lambda_{1} = 69$, $\lambda_{2} = 86$.}
  \label{tab-balanced-sample}      
  \begin{tabular}{r|rrr} \hline 

    & \multicolumn {3}{c}{Predicted}\\ \cline{2-4}

    Classified &{\it CTTS} & {\it WTTS} & {\it FS}\\
    \hline \noalign{\smallskip}

    {\it CTTS} & 29 & 2  & 0 \\
    {\it WTTS} & 3  & 13 & 0\\
    {\it FS}   & 54 & 96 & 7198 \\
    \noalign{\smallskip}\hline

\end{tabular}
\end{table}

\begin{figure}
  \centering
  \includegraphics[width = \columnwidth]{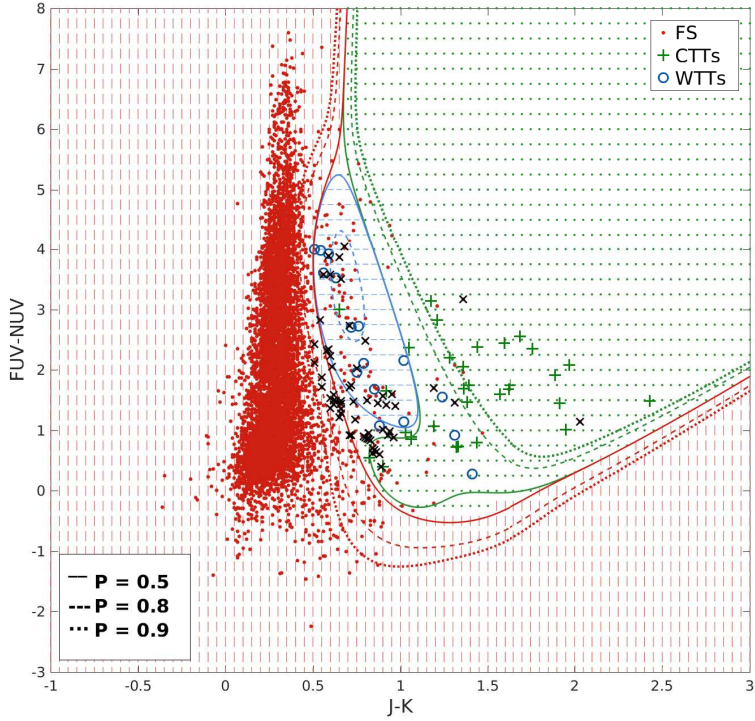}
  \caption{Results of the Logistic Regression fit with a sample balanced using the selected values of
    the scaling coefficients, $\lambda_{1} = 69, \lambda_{2} = 86$. The plane seems to be well
    organised in three different regions assigned to each of the three classes, FS, WTTs and CTTs.
    Some probability lines have been plotted in order to separate the 0.5, 0.8 and 0.9   probability regions.
  Black crosses correspond to the TTs candidates selected by GdC2015.}
\label{fig:tt_lr3}      
\end{figure}

\noindent
The results are shown in Fig \ref{fig:tt_lr3}. 
This time there are some regions of the plane where the membership
  probability to WTTs class is non-zero and greater than 0.5 and the
majority of the WTTs population lies; the slope coefficients
 are listed in Table \ref{tab:lr_tt_best}. With these parameters, we have assigned a
   membership probability to all TTSs candidates\footnote{As a reminder,
     GdC2015 labeled the potential TTS as \textit{Candidate WTTS}, \textit{Candidate CTTS} or just \textit{Candidate} if it was not clear to which class the
     star might belong to. Logistic Regression allows us to assign a membership
     probability to those objects. in GdC2015's Taurus survey to select in the future the best ones for 
follow up observations and subsequent characterisation}; they are
  listed in Table \ref{tab:candidates_probability} and represented in Fig. \ref{fig:tt_lr3} as
  black crosses. It is clear from the plot that there are many FS lying inside
  the $P>0.8$ areas of WTTs and CTTs that were not selected as likely candidates
  to TTs by GdC 2015. We want to highligh that statistical methods are useful
  to derive probability distributions but as intrinsic variations inside
  the sample are not considered, plenty of objects might be misclassified.
  GdC2015 performed a careful examination of the FS sample and even analysed
  the spectral energy distribution curves of some of the sources in order to
  elaborate a reduced but clean sample of TTs candidates. Thus, the combination
of both methodologies is instrumental to achieve reliable results.

\begin{table}
  \centering
  \caption{Slope coefficients of the quadratic Logistic Regression model for the TT balanced sample
    using the set values for the scaling coefficients, $\lambda_{1} = 69, \lambda_{2} = 86$.}
  \label{tab:lr_tt_best}
  \begin{tabular}{ccccccc}
    \hline
    & $\beta_{k,0}$ & $\beta_{k,1}$ & $\beta_{k,2}$ & $\beta_{k,3}$ & $\beta_{k,4}$ & $\beta_{k,5}$ \\
    \hline
    $k=1$   & -10.7245  & 18.1355 & -1.6004 & -6.8158  & -0.1001  & 3.7088 \\
    $k = 2$ & -25.0890  & 38.4523 & 5.8147  & -14.1837 & -0.6552 & -1.7137 \\
    \hline
    
  \end{tabular}
\end{table}

\begin{table}
  \centering
  \caption{Membership probability for TTs candidates($^{a}$) selected by GdC2015 in the Taurus region. The results correspond to the balanced Logistic Regression model
    with $\lambda_{1} = 69$ and $\lambda_{2} = 86$. The full table is available as online material.}
  \label{tab:candidates_probability}
  \begin{tabular}{cccccccccc}
    \hline
    RA & Dec & FUV & NUV & J & K & Type$^{a}$ & P(CTTs) & P(WTTs) & P(FS) \\
    (deg) & (deg) & (mag) & (mag) & (mag) & (mag) & & & & \\
    \hline
    60.176 &   28.871 & 20.21 &  17.87 & 11.17 & 10.58 & Candidate & 0.0808 & 0.5311 & 0.3881 \\
    61.042 &   29.337 & 21.77 &  17.9  &  9.57 &  8.92 & Candidate WTTS &  0.0712 & 0.8435 &   0.0852 \\
    61.498 &   29.944 & 20.85 &  17.68 & 10.19 &  8.83 & Candidate CTTS & 0.9970 & 0.0030 & 0.000 \\
    62.06  &   21.632 & 21.81 &  20.93 &  9.46 &  8.66 & Candidate & 0.4637 &  0.2679  &  0.2684\\
    62.065 &   30.262 & 21.37 &  20.75 & 13.86 & 13.02 & Candidate & 0.5155 &  0.2019  &  0.2826 \\
    62.269 &   28.923 & 21.75 &  18.17 & 9.96  & 9.36  & Candidate WTTS & 0.0415 & 0.7995 & 0.1590 \\
    63.463 &   28.761 & 21.51 &  18.68 & 10.99 & 10.45 & Candidate & 0.0342 & 0.5140 & 0.4518 \\
    64.222 &   26.405 & 16.76 &  16.12 & 13.21 & 12.35 & Candidate & 0.5285 & 0.2357 & 0.2358 \\
    64.716 &   23.858 & 20.9  &  19.19 & 11.29 & 10.58 & Candidate & 0.2441 & 0.5539 & 0.2020 \\
    64.760 &   31.765 & 20.97 &  17.39 & 10.4  & 9.84  & Candidate WTTS & 0.0271 & 0.7075 & 0.2654 \\

    \hline
  \end{tabular}
\end{table}

\section{On the possibility of Main Sequence definition from spectroscopic data}\label{sec-MSclas}

An additional approach, which is worth exploring, is
to define the main sequence (MS) from spectroscopic observations
as has been done for the TTSs. Our qualification
sample has been extracted from the 
high quality photometric catalogue derived by \cite{StarsCatalogue} (hereafter
BAGdC) using IUE spectra.  We selected sources from BAGdC's catalogue
with both FUV and NUV synthetic magnitudes and cross-correlated the results
with the 2MASS catalogue (\cite{Skrutskie2006}) in a search radius of $3''$ to obtain J and K magnitudes;
the spectral types were retrieved from the SIMBAD database (\cite{SIMBAD}). Unfortunately,
only B and A spectral types had a representative
number of sources (we found less than five stars with O, F, G or K 
spectral type); also sdO and sdB populations were numerous enough. Details
of the final sample are listed in Table \ref{tab:ms_sample}.

\begin{table}[h!]
  \caption{IUE sample of MS and sub-dwarf stars with synthetic FUV and NUV
    magnitudes from BAGdC and J and K magnitudes from 2MASS. Spectral types were retrieved
    from SIMBAD.}
  \label{tab:ms_sample}
  \centering
  \begin{tabular}{|ccccc|}
    \hline
    Spectral type  & B,B\texttt{V} & A,A\texttt{V} & sdO & sdB \\
    Stars  & 51 & 26 & 17 & 8 \\
    \hline
  \end{tabular}
\end{table}
  
Due to the poverty of the sample we have not been able to characterise
the MS population as precisely as needed in order to distinguish it from
the young stellar objects.  The details of the calculation can be found
in the Appendix (Sec. \ref{sec-Appendix}).

\section{Conclusions}\label{sec-conclusions}
In this article we have shown that Logistic Regresion,
a mathematical tool seldom used in astronomy, can be
implemented in classification problems. The algorithm has been applied to
assign a probability of being a T Tauri star from 
FUV-NUV versus J-K colour-colour
diagrams. This methodology, combined with a careful
  qualitative analysis of the diagrams (as in GdC2015), provides
  potential candidates to for further scrutiny. Both observational and
  statistical analysis should be carried out together since the method has proven to
  be very dependent on the shape and size of the sample. In this manner, once a sample of candidates has been selected according its properties from colour-colour- diagrams, Logistic Regression can be applied in
order to obtain a membership probability distribution.

\section{Appendix}\label{sec-Appendix}

\begin{figure}
\centering
  \includegraphics[width = \columnwidth]{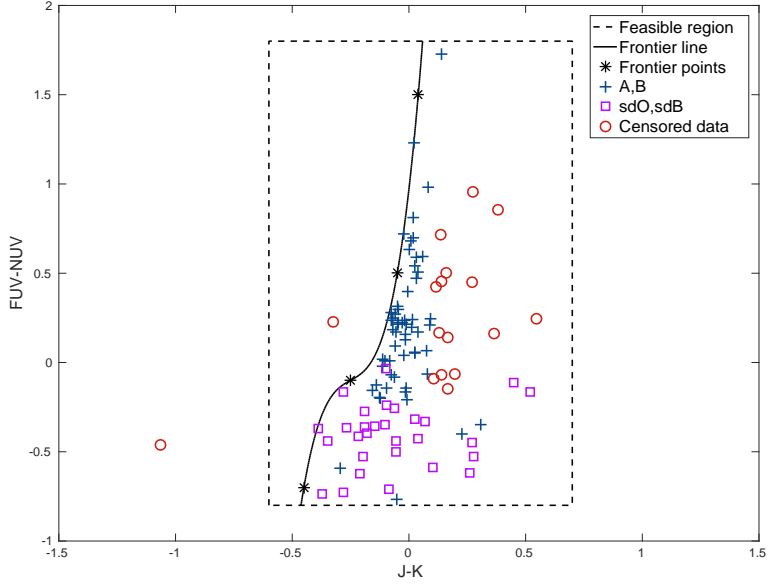}
  \caption{Spatial distribution of the MS stars in the (J-K)-(FUV-NUV) plane. Most of the
    sources lie inside the feasible region and at the right of the predefined cubic curve. MS
    class is represented by $+$ while sub-dwarf class is represented by open squares.
  Censored data are represented with open circles (see text).}
\label{fig:ms_distribution}      
\end{figure}

Provided the poverty of the initial sample of MS stars (see Sec.
\ref{sec-MSclas}), we decided to 
combine the four clases into two big groups: MS stars (B and A classes)
and sub-dwarfs (sdO and sdB); the spatial distribution of the sources
in the $x_{1}=$J-K vs $x_{2}=$FUV-NUV plane is shown in Fig. \ref{fig:ms_distribution}.
In a natural way, the
stars are confined in a subregion of the whole plane $\mathbb{R}^{2}$, so it is
coherent to define a smaller area or feasible region that is expected to 
contain the vast majority of the sources and that facilitates
the obtention of more accurate results; in this case, it has been defined
as $-0.6 \leq {\rm J-K} \leq 0.7$ and $-0.8 \leq {\rm FUV-NUV} \leq 1.8$. This subspace has been
further divided by adjusting the sequence of A and B stars following a third
degree equation that passes through four simulated points
(see Fig. \ref{fig:ms_distribution}): 
$ x_2 = 0.97 + 11.42 \cdot x_1 + 42.81 \cdot x_1^2 + 57.08 \cdot x_1^3 $. Once
all these assumptions were made, we dropped from the sample the two points that were
out of the feasible region as well as other MS stars that showed abnormally infrared
excesses.

\subsection{Logistic Regression - Application to two classes}\label{subsec-MS_LR}

In this case, the classification problem
consists on distinguishing between two classes ($K = 2)$: 
\textit{AB} ($k = 1$) and \textit{sd} ($k = 2$) . Therefore,
given one of these objects \textit{O} the problem reduces
to determine the probability of belonging to the first class, \textit{AB}:

\begin{equation}
  P(AB|\textbf{x}^{o}) = \frac{exp(f(x))}{1+exp(f(x))} \doteq p_{1}
  \label{eq:p1}
\end{equation}

and so, the probability of membership to the \textit{sd} class is:

\begin{equation}
  P(sd|\textbf{x}^{o}) = 1-P(AB|\textbf{x}^{o}) = 1-p_{1} = \frac{1}{1+exp(f(x))} \doteq p_{2}
  \label{eq:p2}
\end{equation}

The independent variables of the regression model are in a natural
way $x_{1} = J-K$ and $x_{2} = FUV-NUV$. It is clear in Fig. \ref{fig:ms_distribution} that the
classification cannot be linear and so, after some computational
experience, the variables have been extended to $x_{3} = x_{1}^{2}$, $x_{4}=x_{2}^{2}$ and
$x_{5} = x_{1}x_{2}$ to
perform a quadratic Logistic Regression. Thus, the linear combination
of the predictor variables is given by
\begin{align}
  f(x) = \beta _{1,0} + \beta _{1,1}x_{1} + \beta _{1,2}x_{2} + \beta_{3}x_{3} + \beta_{1,4}x_{4} + \beta_{1,5}x_{5} =  \\
  \beta _{1,0} + \beta _{1,1}x_{1} + \beta _{1,2}x_{2} + \beta_{1,3}x_{1}^{2} + \beta_{1,4}x_{2}^{2} + \beta_{1,5}x_{1}x_{2} \nonumber
  \label{eq:f1}
\end{align}

The parameters are listed in Table \ref{tab:coef_ms}.

\begin{table}[h!]
  \centering
  \caption{Slope coefficients of the quadratic Logistic Regression
  model for the MS sample, that corresponds to the \textit{AB} class ($k =1$). }
  \label{tab:coef_ms}
  \begin{tabular}{cccccc}
    \hline
    $\beta_{1,0}$ & $\beta_{1,1}$ & $\beta_{1,2}$ & $\beta_{1,3}$ & $\beta_{1,4}$ & $\beta_{1,5}$ \\
    \hline
     1.9343 & 1.3240 & 9.9596 & -14.9457 & 9.9011 & -0.9324 \\
    \hline 
  \end{tabular}
\end{table}

\begin{figure}
\centering
  \includegraphics[width = \columnwidth]{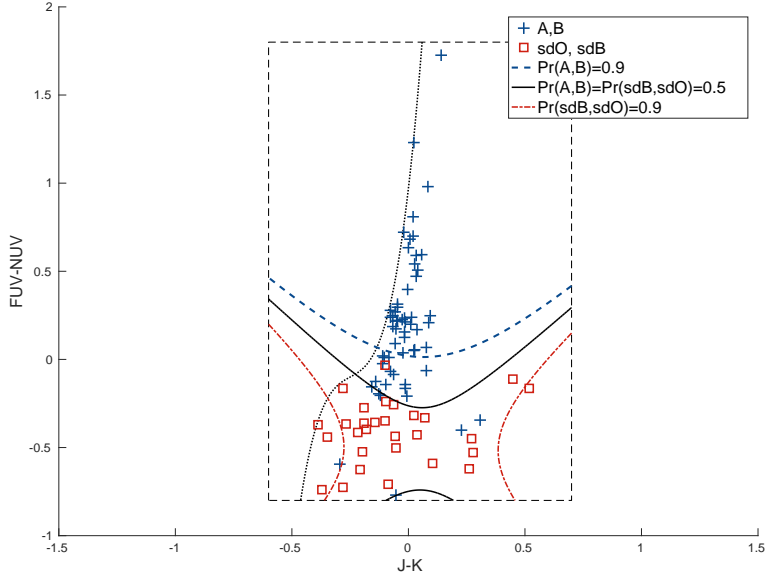}
  \caption{Probability lines for the MS sample. The solid black line
    indicates de region of equal probability between \textit{AB} and
    \textit{sd} classes; the dashed blue (dashed-dotted red) line
    limits the region of 0.9
    probability of belonging to the \textit{AB} (\textit{sd}) class.}
\label{fig:ms_probability}      
\end{figure}

Using these results, we have plotted as a guidance 
the probability lines for
$p_{1} = 0.5$, $p_{1} = 0.9$ and $p_{1} = 0.1$ over the  original
data (Fig. \ref{fig:ms_probability}). The two populations are fairly separated and more than half of the \textit{AB} points lie inside the 0.9 probability region.

The object classification in this framework is made in a simple and 
natural way: an object \textit{O} belongs to the \textit{AB}
class if $P(O \in AB|\textbf{x}^{o}) > P(O \in sd|\textbf{x}^{o})$, where the
probabilities are given by Eqs. \ref{eq:p1}, \ref{eq:p2}. The ideal situation would be that where both
probabilities are the same, that is to say, $p_{1} = 0.5$;  the confusion matrix is shown in Table
\ref{tab:conf_05}.

\begin{table}[h!]
  \caption{Confusion matrix for \textit{MS} and \textit{sd} stars, taking  $p_{1} = 0.5$.}
  \label{tab:conf_05}
   \centering
    \begin{tabular}{r|rr} \hline 

    & \multicolumn {2}{c}{Predicted}\\ \cline{2-3}

    Classified &{\it AB} & {\it sd} \\
    \hline \noalign{\smallskip}

    {\it AB} & 55 & 3   \\
    {\it sd} & 2  & 27 \\
    \noalign{\smallskip}\hline
    \end{tabular} \\
\end{table}

\begin{figure}
\centering
  \includegraphics[width = \columnwidth]{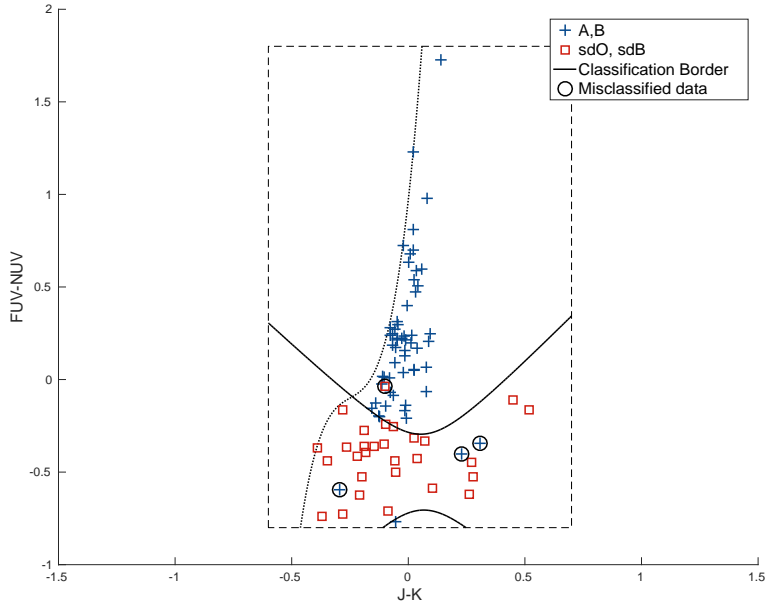}
  \caption{Best classification of the MS population after applying the Logistic
    Regression. The line that separates the two classes corresponds to a probability
    of belonging to the \textit{AB} class of $p_{1} = 0.474$ with a 4.6\% probability
  of misclassification.}
\label{fig:ms_classification}      
\end{figure}

\begin{figure}
\centering
  \includegraphics[width = \columnwidth]{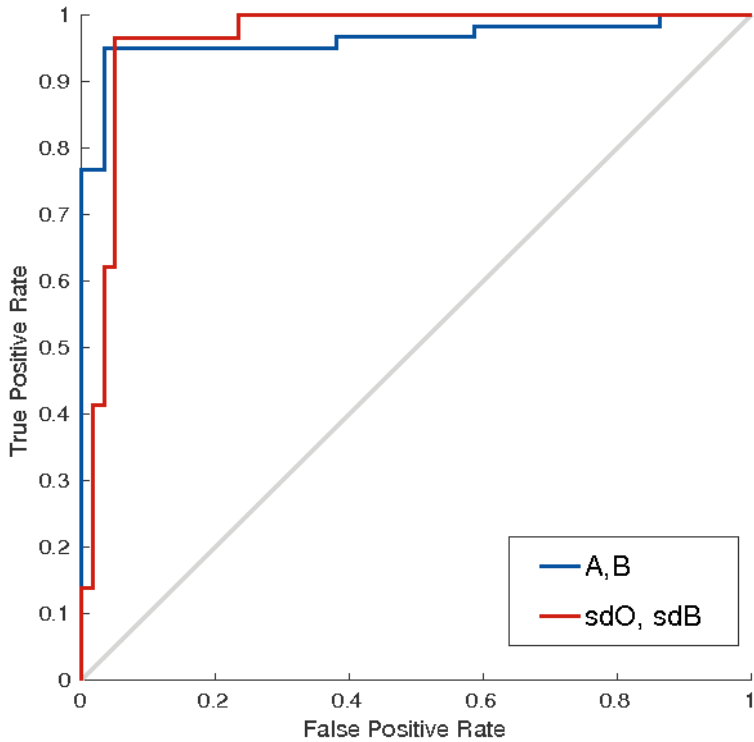}
  \caption{ROC curve for the final classification of the MS population (TPR vs. FPR), considering $P(AB|\textbf{x}^{o}) = 0.474$. The
  blue line corresponds to the \textit{AB} class, whereas the green line corresponds to the \textit{sd} class.}
\label{fig:roc}      
\end{figure}

Nevertheless, the variable nature of the data suggests
that a better classification may be achieved, in
the sense that the confusion matrix could be improved to get a minimum
number of misclassified objects. We have tested several values
for the $p_{1}$ parameter, among which are $p_{1}=0.487$, $p_{1} = 0.474$ and $p_{1} = 0.462$ (see
their confusion matrices in Table \ref{tab:three-matrices})
and found that an acceptable result is achieved with $p_{1} = 0.474$.
Graphically (see Fig. \ref{fig:ms_classification}), the vast majority of the data are well classified except for
three B stars, PG 1210+429, BD-13 4920 and CSI+81-09137, that lie among the \textit{sd} 
population and one sdO star, PG 2219+094, located among the \text{AB}
population. The quality of this classification is also supported by the excellent
results obtained for the area under the \textit{Receiver Operating Characteristic} (ROC) 
curve (see Fig. \ref{fig:roc}), \textit{AUC}, for the positive condition for
MS and sd classes; in both cases $AUC = 0.9645$. ROC curve represents the probability 
of detecting the true condition versus the probability of detecting the false
condition , and thus it is restricted to the plane
$[0,1]\times[0,1]$. AUC has become
a standard for evaluating classification procedures; the greater the values
under the curve, the better the classification.


\begin{table}[h!]
  \caption{Confusion matrices for several values of $p_{1}$.}
  \label{tab:three-matrices}
  \begin{minipage}{0.32\textwidth} 
    \centering
    \begin{tabular}{r|rr} \hline 

    & \multicolumn {2}{c}{Pred.}\\ \cline{2-3}

    Clas. &{\it AB} & {\it sd} \\
    \hline \noalign{\smallskip}

    {\it AB} & 53 & 5   \\
    {\it sd} & 1  & 28 \\
    \noalign{\smallskip}\hline
    \end{tabular} \\
    $p_{1} = 0.487$
  \end{minipage}
  \begin{minipage}{0.32\textwidth} 
    \centering
    \begin{tabular}{r|rr} \hline 

    & \multicolumn {2}{c}{Pred.}\\ \cline{2-3}

    Clas. &{\it AB} & {\it sd} \\
    \hline \noalign{\smallskip}

    {\it AB} & 55 & 3   \\
    {\it sd} & 1  & 28 \\
    \noalign{\smallskip}\hline
    \end{tabular} \\
    $p_{1} = 0.474$
  \end{minipage}
  \begin{minipage}{0.32\textwidth} 
    \centering
    \begin{tabular}{r|rr} \hline 

    & \multicolumn {2}{c}{Pred.}\\ \cline{2-3}

    Clas. &{\it AB} & {\it sd} \\
    \hline \noalign{\smallskip}

    {\it AB} & 55 & 3   \\
    {\it sd} & 2  & 27 \\
    \noalign{\smallskip}\hline
    \end{tabular} \\
    $p_{1} = 0.462$
  \end{minipage}

\end{table}

To sum up, an object of the MS population is classified as part of the
\textit{AB} class if $p_{1} \geq 0.474$, and
of the \textit{sd} class in another case, with a 4.6\% probability of misclassification.


\begin{acknowledgements}
We would like to thank the referee for her/his useful comments that have improved the article.
\end{acknowledgements}

\bibliographystyle{spbasic}      
\bibliography{references}   

%
%

\end{document}